\documentclass{aastex62}

\accepted{August 14, 2019}
\submitjournal{ApJ}

\shorttitle{Superluminal Motion in GRBs}
\shortauthors{Hakkila \& Nemiroff}

\begin{document}

\title{Time-Reversed Gamma-Ray Burst Light Curve Characteristics \\as Transitions between Subluminal and Superluminal Motion}

\correspondingauthor{Jon Hakkila}
\email{hakkilaj@cofc.edu}

\author{Jon Hakkila}
\affiliation{The Graduate School, University of Charleston, SC at the College of Charleston, 66 George St., Charleston, SC 29424-0001, USA}
\affiliation{Department of Physics and Astronomy, College of Charleston, 66 George St. Charleston, SC 29424-0001, USA}

\author{Robert Nemiroff}
\affiliation{Department of Physics and Astronomy, Michigan Technological University}




\begin{abstract}

We introduce a simple model to explain the time-reversed and stretched residuals
in gamma-ray burst (GRB) pulse light curves. In this model
an impactor wave in an expanding GRB jet accelerates
from subluminal to superluminal velocities, or 
decelerates from superluminal to subluminal velocities.
The impactor wave interacts with the surrounding medium
to produce Cherenkov and/or other collisional
radiation when traveling faster than
the speed of light in this medium, and other mechanisms (such as thermalized Compton
or synchrotron shock radiation) when traveling slower than
the speed of light.
These transitions create both a time-forward and a time-reversed
set of light curve features through the process of Relativistic Image Doubling (RID).
The model can account for a variety of unexplained yet observed GRB pulse
behaviors including the amount of stretching observed in 
time-reversed GRB pulse residuals and the relationship between 
stretching factor and pulse asymmetry. 
The model is applicable to all GRB classes since similar pulse
behaviors are observed in long/intermediate GRBs, short GRBs, and x-ray flares.
The free model parameters are the impactor's Lorentz factor
when moving subluminally, its Lorentz factor when moving superluminally, 
and the speed of light in the impacted medium.


\end{abstract}




\section{Introduction} \label{sec:intro}

Gamma-ray bursts (GRBs) originate from the violent ejection of material 
at the end of massive star evolution accompanying black hole formation 
\citep{woo93, hjo03, sta03,wb06}. 
This ejection takes the form of collimated relativistic outflows \citep{saa99}
that must blast through a star's envelope before producing the observed gamma-rays 
\citep{mac99, alo00}. Prompt GRB emission is produced
sometime after a jet develops and before the afterglow phase.
The afterglow is produced when the 
relativistic outflow slows while sweeping up and 
accelerating material in an external medium; this external medium
radiates via synchrotron radiation ({\em e.g.}, \cite{mes97,pan98,sar98,wax98,gro99,wij99,che00,pan00,kob00}).

In recent years, GRB models have become increasingly complex.
For example, adaptive mesh refinement codes have been developed to 
simulate the GRB jet ejection process \citep{mor07, laz09, laz10, mor10, laz11, nag11}.
Two-dimensional \citep{mac99,alo00,mac01,zha03, miz06, mor07, 
mor10, laz09, laz10, laz11, miz09,nag11} and more recently three-dimensional simulations 
({\em e.g.,} \cite{zha04,lopez13}) have improved our understanding of
GRB jet development. These models have shown that particles in GRB jets are accelerated
in the vicinity of the developing accretion disk and again upon exiting the stellar photosphere.

Although computational models have
helped improve our understanding of the conditions 
resulting in GRB energetics, they have not been able
to provide key answers about the processes that produce GRB prompt emission. 
Neither the events responsible for producing variations in GRB light curves
nor the mechanism(s) responsible for producing GRB spectra are
clearly understood, and computational models are only now 
reaching the point where they might be constrained by these observations.

\subsection{Characteristics of GRB Pulses}

Central to our understanding of GRB light curves are the observations that
their behaviors are characterized by generic and repeatable patterns that likely indicate kinematics
within the jet. 
Measurement and characterization 
of light curve properties, and identification of these patterns, has been impeded by the low signal-to-noise-ratio regime
in which they are typically observed. However, once signal-to-noise ratio and other instrumental properties
have been accounted for, GRB prompt emission exhibits a number of recognizable characteristics:
\begin{itemize}
\item {\em Pulses} lasting from milliseconds to hundreds of seconds form the basic units of GRB prompt emission \citep{nor96}, and pulses exhibiting similar characteristics constitute most of the emission from long/intermediate GRBs ({\em e.g.} \cite{nor86,nor96,pen97,nor05,hak18b}), short GRBs ({\em e.g.} \cite{nor11,hak14,hak18a}), and x-ray flares found in GRB afterglows ({\em e.g.} \cite{mar10,hak16}).
\item Pulses can be modeled to first order by simple asymmetric monotonic functions ({\em e.g.,} \cite{nor96,ste96,lee00a,lee00b,koc03,nor05,nem12,bha12}), with pulse decay times generally being longer than pulse rise times.
\item Pulses evolve from hard-to-soft, with asymmetric pulses showing more pronounced spectral evolution than symmetric pulses ({\em e.g.} \cite{cri99,ryd99,hak11,hak15}). When early pulse emission can be observed, pulses are shown to start near-simultaneously at all energies \citep{hak09}.
\item Pulses are not defined solely by monotonic emission, but also by a set of {\em residual} structures that overlay the monotonic pulse component \citep{bor01,hak14}. These residual structures can be found by subtracting the monotonic pulse component from the light curve. The residuals exhibit considerable complexity at high signal-to-noise ratios, but much of this complexity washes out and the residuals take the form of a triple-peaked wave at moderate- signal-to-noise ratios \citep{hak18a,hak18b}. This triple-peaked structure deviates from monotonicity during the pulse rise (the {\em precursor peak}), near the pulse maximum (the {\em central peak}), and during the pulse decay (the {\em decay peak}).  At low signal-to-noise ratios only monotonic pulse components can be reliably recovered.
\item These residual structures do not constitute unrelated events, but are instead causally linked to each other as well as to the underlying monotonic pulse. The causal linkage occurs because each structure can be folded at a {\em time of reflection}, then stretched until the time-reversed residuals at the beginning of the pulse match those at the end of the pulse \citep{hak18b}. The asymmetry of the residual structure is directly related to that of the monotonic pulse. The time of reflection for a time-reversible structure usually (but does not always) coincides with the time of monotonic component's peak intensity \citep{hak18a,hak18b}. 
\item Broadband time-evolving spectral analyses are often used to describe GRB spectral evolution because of poor photon counting statistics. These analyses have identified ``hard-to-soft'' and ``intensity tracking'' behaviors ({\em e.g.,} \cite{whe73, gol83, nor86, pac92}), but this oversimplified bimodal classification scheme has been shown to represent a continuous range of behaviors ({\em e.g.,} \cite{kar94, bha94, for95,bor01}). Treating each GRB pulse as a {\em structured} episode rather than one in which pulses are statistically-significant intensity peaks, several studies \citep{lia96,hak15,hak18a,hak18b} have demonstrated that most pulses undergo hard-to-soft spectral evolution in that they are harder at the pulse onset than they are late in the decay phase. This evolution fluctuates rather than being smooth and gradual because GRB pulse spectra re-harden around the times that they re-brighten. Pulses appear to evolve hard-to-soft if their central peaks are softer than their precursor peaks, whereas they follow intensity tracking behaviors if their central peaks are harder than their precursor peaks. Depending on the relative hardnesses of the peaks, there exists a wide range of intermediary behaviors \citep{hak15} bounding these extremes. 
\item Pulse asymmetry correlates with hardness evolution: asymmetric pulses are harder overall and have more pronounced hard-to-soft evolution than symmetric pulses \cite{hak15}, which more commonly exhibit ``intensity tracking'' behaviors. Structured features, or ``spikes" representing strong variability in GRB pulses are generally harder than the underlying smooth pulse emission. These structures generally can be characterized as belonging to the time-reversed GRB pulse light curve component.
\item A few time-dependent pulse spectral studies performed on bright GRBs find that the pulse rise spectrum often differs spectrally from that characterizing the pulse decay. Specifically, some early hard photon production is consistent with thermal heating in an optically-thick photospheric region ({\em e.g.,} \cite{ryd09,gui11,zha11,ryd11,yu2015, peer18,lu18}). This interpretation might help explain in part why GRB spectra are inconsistent with the synchrotron shock model ({\em e.g.} \cite{tav95}); this model does not produce a sufficient number of low-energy photons to match the data \citep{pre98}. These time-dependent studies indicate that one simple combination of spectral components is not capable of explaining all GRB spectra, which is consistent with the broadband results indicating that the spectrum may have distinct episodes in which additional energy is injected into the spectrum.
\end{itemize}

\begin{figure}
\plottwo{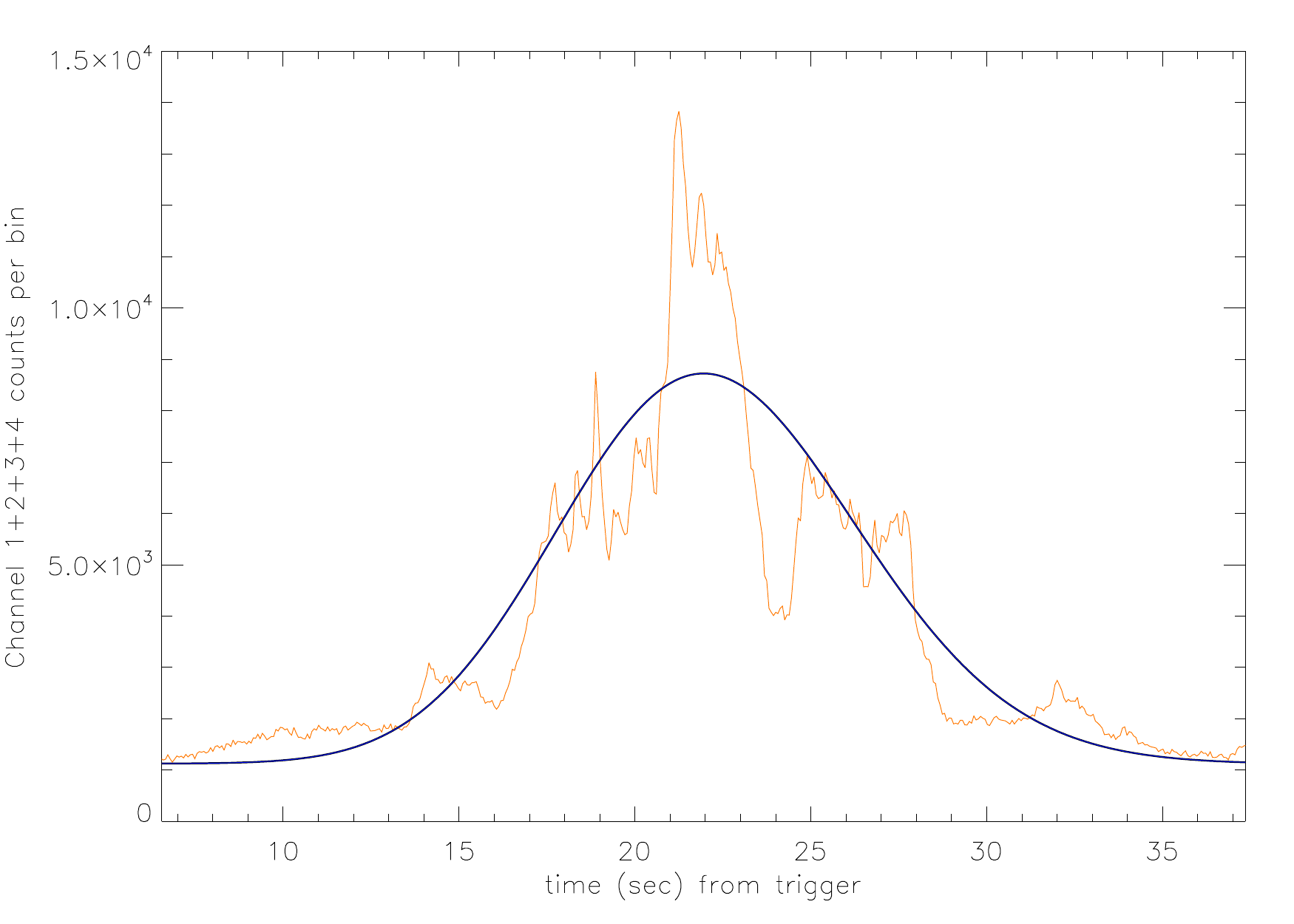}{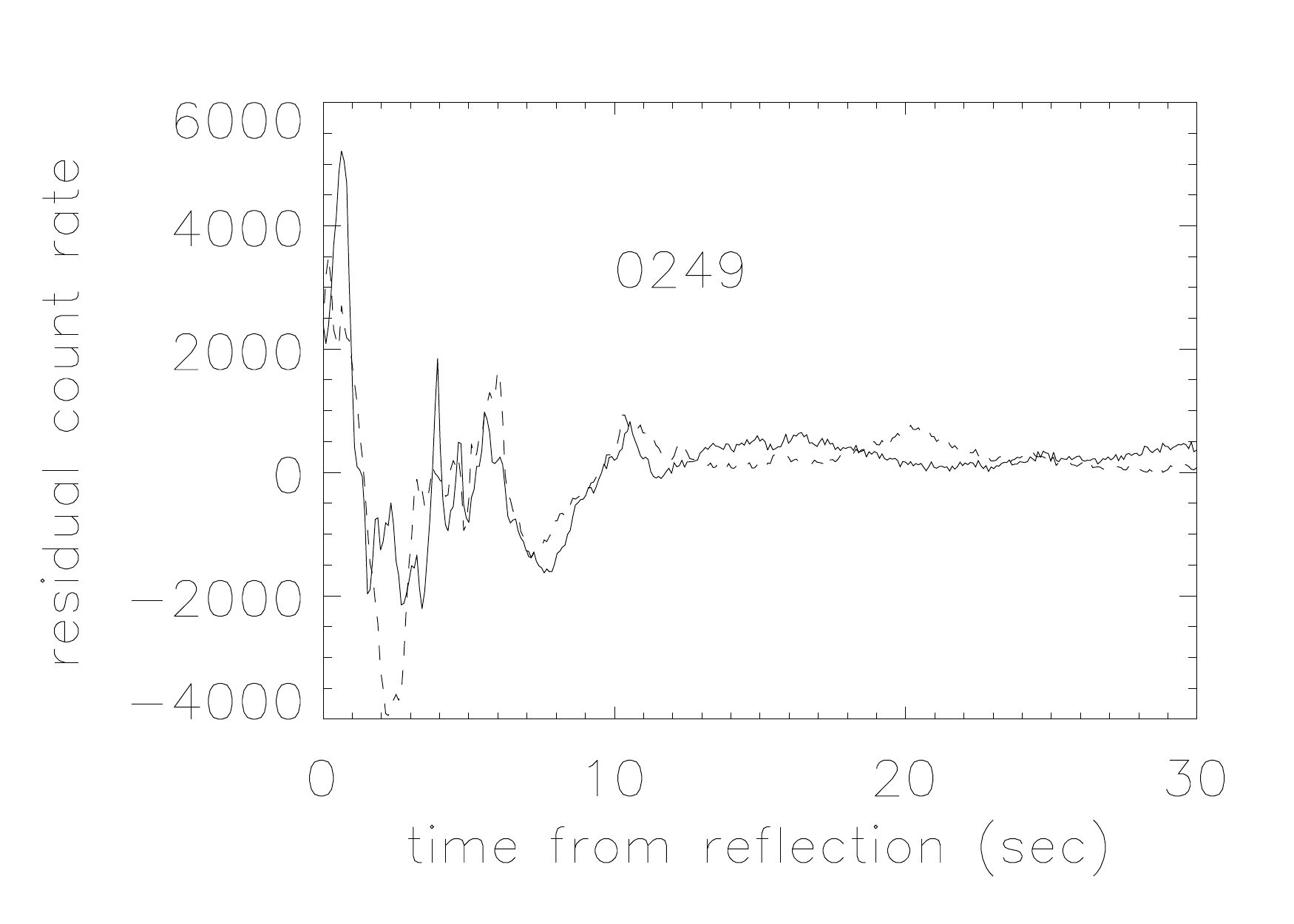}
\caption{The four-channel light curve of single-pulsed BATSE GRB 249 (left panel), showing the \cite{nor05} pulse characterization of the monotonic pulse component (dashed line). Also shown (right panel) are the light curve residuals, folded and stretched at the time of reflection to show how well they match one another (the solid line represents the folded and stretched residuals preceding the time of reflection, while the dashed line represents the residuals following that time). The process used to identify time-reversality, as well as the fitted characteristics of this pulse, is described in \cite{hak18b}. \label{fig:249}}\end{figure}

The aforementioned observed characteristics have improved our ability to characterize GRB pulses: {\em Whereas a GRB light curve once seemed to comprise a series of unrelated chaotic short-duration events, this same light curve now appears to contain only a few longer-duration pulses which can each be identified by their time-reversible structures.} Figure \ref{fig:249} demonstrates time-reversed and stretched characteristics of the complex single-pulsed BATSE trigger 249 ($\kappa=0.07, s_{\rm mirror} = 0.74$; {defined in Section \ref{sec:accel}).} Most, but not all GRBs can be characterized in this fashion; we cautiously note that the common pulse behaviors described here may not be universal.

\subsection{Towards Explaining GRB Pulses}

Under the right conditions, kinematic motions can provide a framework for
explaining the time-reversed and stretched light curve components found in GRB pulses.
\cite{hak18b} proposed several simple geometric GRB jet models
capable of producing residuals with these characteristics.
However, each model contained at least one {\em ad hoc} assumption
about the interaction between a single {\em impactor} (a soliton or single clump of
particles ejected along the jet path by the central engine) and clumps within the jet.
One model hypothesized the existence of a physical barrier (such as the jet head)
to reflect and reverse the impactor's motion as the jet catches up with it. Another model
required the impactor to move through a bilaterally-symmetric distribution
of material within the jet. A third model incorporated the bilaterally-symmetric
distribution into the impactor which then collided with a single jet structure.
Yet other models required a bilaterally-symmetric wave to form by itself within the jet.
Each of these models required some sort
of unexplained kinematic symmetry, either in a forward/reverse motion of the impactor or 
by a bilaterally symmetric structure of the medium or the impactor.
None of these models seemed to be directly connected to other inferred
characteristics of GRB jets. 

Few attempts have been made to incorporate observed pulse properties
into adaptive mesh jet models. A few recent analyses \citep{par18a, par18b}
have tried incorporating radiative transfer into jet models, but
these models have not addressed GRB pulse properties. {\em A bridge needs to
be built between observed properties of GRB pulsed emission and 
realistic GRB computational models. We propose to incorporate a model explaining
the characteristics of GRB pulse light curves
in an attempt to span this gap.}

In this manuscript we explore an odd natural property that has the
potential of producing time-reversed and stretched,
hard-to-soft, structured light curves without
relying on either kinematic or structural symmetry.
This property is superluminal motion
and the process of Relativistic Image Doubling, or RID ({\em e.g.,} \cite{nem18}), which can
occur when a relativistic emitter or impactor's velocity exceeds the 
speed of light in the medium through which it travels. 
Given the large Lorentz factors ($\Gamma=100$ or higher) inferred from GRB observations
and the lack of study performed in this area,
this possibility is not as unreasonable as it first sounds. 
In accounting for GRB pulse characteristics, we seek to retain consistency 
with other GRB observations and with standard features
of adaptive mesh GRB kinematic models. We also hope that
our explanation will prove to be consistent with GRB spectral
properties and their associated theoretical models.

\section{Model Basics: Time-Reversed Light Curves and Superluminal Motion} \label{sec:super}

\begin{figure}
\epsscale{0.75}
\plotone{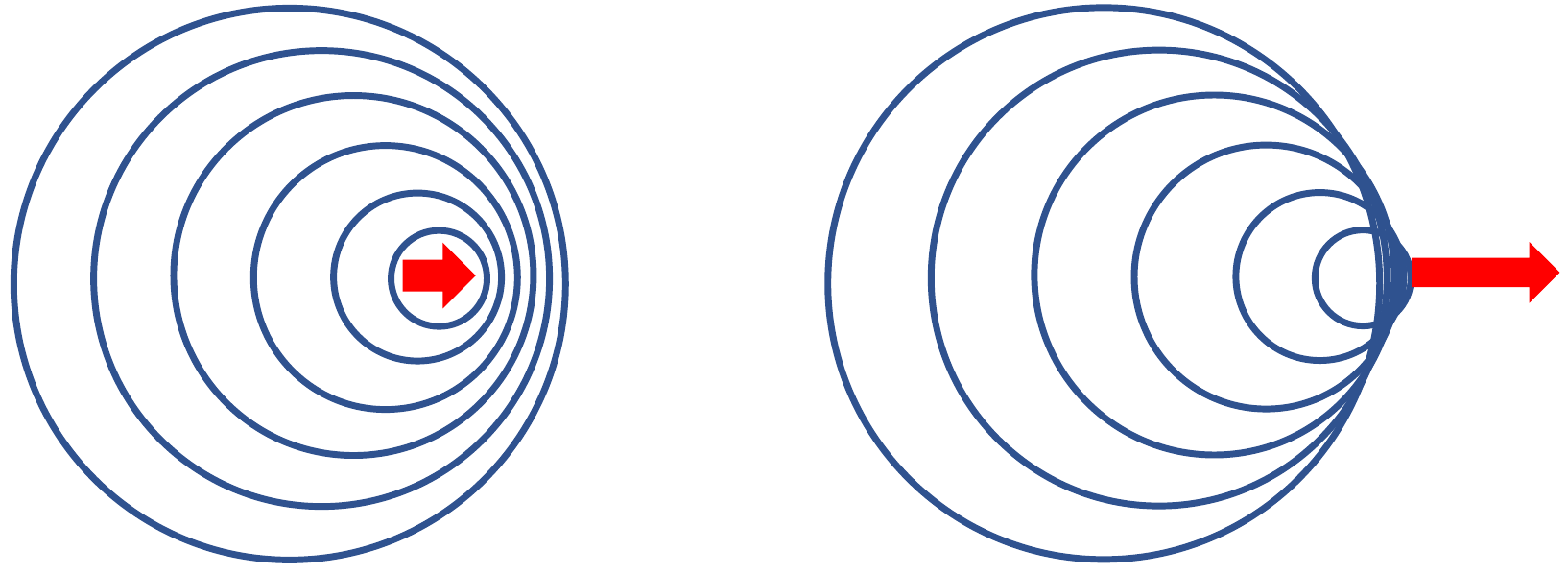}
\caption{Light emitted by a subluminal source (left) as it accelerates to become a superluminal source (right). \label{fig:accelerate}}\end{figure}

Under the right conditions, charged particles moving superluminally through a polarizable medium can induce 
the medium to emit Cherenkov radiation. Cherenkov radiation \citep{che34,tam37} is thus a form of 
continuous radiation that derives from a 
medium and not from the superluminal particles passing through it. 
To a stationary observer,
a superluminal impactor can move in advance
of its own spherically expanding emission; this mechanism contrasts with a
subluminal emitter that is surrounded by its expanding light sphere
(see the left panel of Figure \ref{fig:accelerate}).
The induced emission from a superluminal impactor
(see the right panel of Figure \ref{fig:accelerate})
radiates isotropically but is seen by a stationary forward observer
as waves of trailing emission
(sometimes following a bright flash) moving toward the observer as a result of timing and geometry. 
If the observer is directly in front
of the approaching emitter, the impactor's most recent emission
should be followed by progressively earlier emissions.
However, the observer cannot see this time-reversed
radiation until the impactor slows to subluminal velocities,
at which time the most recently-emitted light terminates
the superluminal chain of emission.
These characteristics form the basis for RID \citep{nem18}.

A superluminal impactor can travel through a medium at speeds
exceeding the speed of light in that medium. 
When the impactor eventually slows as it emerges from this medium at a later time,
the radiation produced behind it during its superluminal motion continues to follow it in reverse order.
An emitter of this type can play its superluminal
history in time-reversed order as it approaches an observer.

Citing MHD models such as \cite{lopez13}, we consider an ejection of material driven from its progenitor star that rapidly
develops into a relativistic jet with a typical Lorentz factor of $\Gamma = 100$. Emissions from the jet must point almost directly towards an observer in radial direction $r$ in order to be seen, while a GRB pulse observed to have duration $\tau = 10s$ must have a source frame duration of $\tau_0 = \tau / \Gamma \approx 0.1s$. Now consider an impactor expanding outward from the progenitor along the jet axis; this may either be a disturbance related to the head of the jet as it expands into the surrounding circumstellar medium or an instability moving through the expanding jet material.

For our purposes, we propose that impactor is not a soliton or single blast wave as described by \cite{hak18b}, but that it instead takes the form of an {\em impactor wave} capable of producing half of the wavelike structure observed in a GRB pulse light curve, once a fitted monotonic pulse has been removed. This single wavelike feature is responsible for producing both the `time forward' and `time reversed' pulse residuals. Sample residual wave characteristics are shown for BATSE trigger 249 in the right panel of Figure \ref{fig:249}. The maximum amplitudes of this residual structure occur toward the pulse center where the wavelength is shortest, while the minimum wave amplitudes occur in the wings where the wavelength increases. We propose that the material ejected from the central engine either already has this wave property or develops it as it moves. The wave might represent density/pressure variations, or it could indicate local abrupt changes in the internal magnetic field. Such variations might be the result of a plasma instability but could also capture the memory of conditions near the central engine as the jet was being produced \citep{kob97,dai98,pan99}. The abrupt termination/initialization of the wave could coincide with the head of the jet.

In the inner jet (close to the progenitor) during the jet breakout process, the velocity of the impactor wave $v_o$ is less than the speed of light in the medium $c_J$. In other words, $v_o < c_J < c$ (where $c$ is the speed of light in vacuum). Acceleration of the jetted material occurs as the jet opens at the stellar surface when thermal energy is converted to kinetic energy ({\em e.g.,} \cite{miz13,lopez13}). {\em We propose that the impactor wave velocity in the central portion of the jet $v_i$ now exceeds the speed of light in the jet medium ($v_i > c_J$);} it remains superluminal until the impactor slows and returns to subluminal speeds far from the progenitor. If valid, this assumption allows either one of the two transitions between subluminal and superluminal impactor motion to produce time-reversed and stretched GRB pulse light curves. The first of these transitions occurs when the impactor accelerates from subluminal to superluminal speeds during the jet breakout phase and leads to our ``Impactor Acceleration" model. The second occurs when the impactor decelerates from superluminal to subluminal speeds as the jet sweeps up enough interstellar medium to slow, just before the beginning of the afterglow phase (the ``Impactor Deceleration" model). To simplify our models, we consider the speed of light in the medium to be essentially constant over a range of impactor velocities. In reality, we recognize that an impactor can also become superluminal at constant velocity if the speed of light in the medium decreases ({\em e.g., } via ionization, magnetization, or other effects) during the impactor's passage. 

The superluminal velocity of the wave causes its early emissions to lag behind its later emissions, starting at the moment it drops below superluminal velocities. The second half of the residual wave produced by this mechanism is observed to be a mirrored series of events to those associated with the wave's first half.

\section{Impactor Wave Acceleration} \label{sec:accel}

\begin{figure}
\plotone{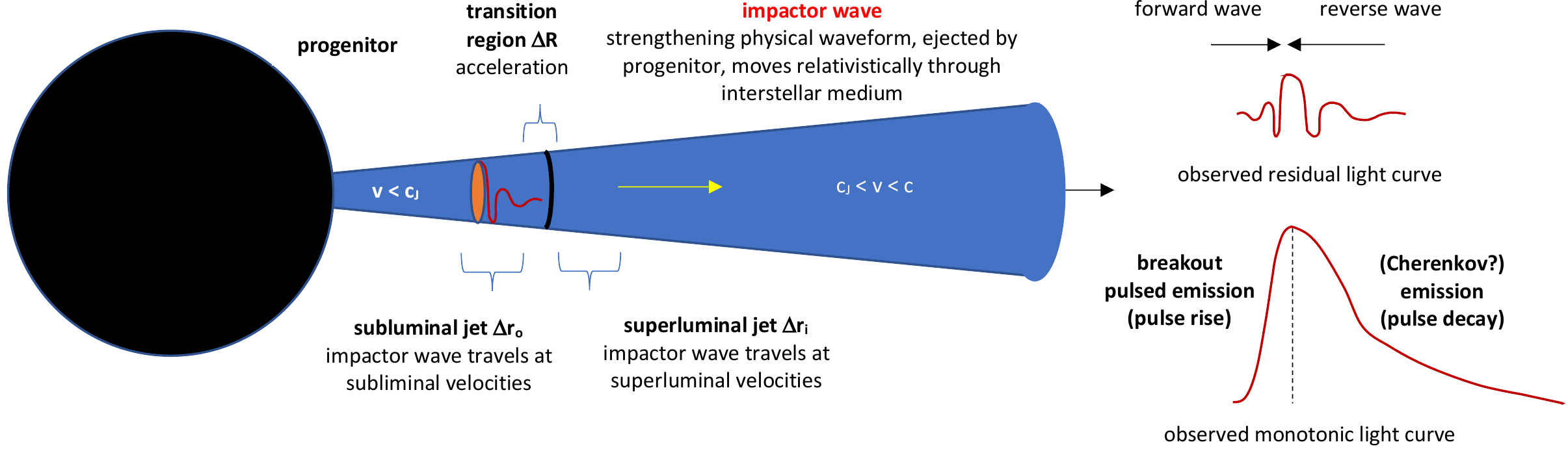}
\caption{The impactor Acceleration Model. Within a GRB jet, an impactor wave accelerates from subluminal to superluminal velocities in the inner part of the jet. The impactor wave increases in amplitude and terminates upon reaching its maximum value. \label{fig:model}}\end{figure}

In the Impactor Acceleration model, the relativistic motion of the impactor wave towards the observer produces a correctly ordered and time-compressed (Doppler shifted) series of events, while its subsequent superluminal motion produces a time-reversed and stretched image of the same events. A model of this is shown in Figure \ref{fig:model}.

The perceived approach velocity of the impactor wave generating light in a relativistic jet is given in Equation 1 of \cite{nem18}:
\begin{equation}
u_{approach} = \frac{v}{(1-v/c)}
\end{equation}
where $v$ is the wave's velocity and $c$ is the speed of light. Note that when $v < c$ this velocity is positive towards the observer but when $v > c$ this velocity will be negative such that the approaching source will appear to move away from the observer.

We assume that the impactor wave generates radiation as it passes subliminally outward through a region of dimension $\Delta r_o$. Although we do not know the mechanism that might produce this radiation, we suggest that conditions preceding the subluminal to superluminal transition may play a part. Exceeding a medium's speed of light drains kinetic energy from an impactor \citep{tam60} and thus may supply collisional and radiative energy to the system. Our definition merely states that the inner boundary of $\Delta r_o$ identifies the jet location at which subluminal radiation begins, while the outer boundary indicates the location at which this radiation ends.
Similarly, we assume that the impactor generates radiation primarily by Cherenkov radiation (but perhaps also via collisions) as it passes superluminally through a region of radial dimension $\Delta r_i$. The transition from subluminal to superluminal motion occurs in a region of radial dimension $\Delta R$. All of the radiation is produced in the jet region $\Delta r_o + \Delta R + \Delta r_i$.

The perceived approach velocity of the subluminal impactor is
\begin{equation}
u_o = \frac{v_o}{(1-v_o/c_J)},  \> \> \> \> \> \> v_o < c_J < c
\end{equation}
whereas its perceived approach velocity after it has gone superluminal is
\begin{equation}
u_i = \frac{v_i}{(1-v_i/c_J)}, \> \> \> \> \> \> c_J  < v_i < c.
\end{equation}

We assume that radiation associated with the residuals is produced only when the impactor wave of length $R_{\rm wave}$ passes through $\Delta r_o$ and $\Delta r_i$ (defined in the observer's frame). Thus, each of these regions cannot be too large in order to produce proportional wave-like residuals, for if they extended to radial dimensions similar to those of the wave-like impactor then the wave structure would be smeared out by simultaneous photon production from different locations within the emitting regions. Similarly, the wave must pass entirely through $\Delta r_o$ before it begins to reverse in $\Delta r_i$, or else the light from subluminal structures would overlap some of those produced superluminally and the time-reversed residual structure would not be obvious to an observer. As indicated previously, the impactor wave must initially have a small amplitude that gets larger until it crescendos and terminates in order to explain the observed residual wave characteristics. Physically, this structure might represent oscillations produced near the progenitor at the beginning of the outflow that increase until they terminate with an explosive outflow.

$\Delta R$ must be larger than the physical extent of the impactor because $\Delta R$ must be large enough that the residual light curve produced by the impactor wave, traveling within the jet at speed $c_J$, passes the entirely of the impactor wave, moving at speed $v_o$. The time it takes for this to occur is $t=R_{\rm wave}/(c_J-v_o)$. During this time, the impactor wave travels a distance within the jet $\Delta R=c_J t = R_{\rm wave}/(1-v_o/c_J)$. When $v_o \approx c_J$, $\Delta R >> R_{\rm wave}$; it is only in the limit of $v_o << c_J$ that $\Delta R \rightarrow R_{\rm wave}$. In other words, a rapid transition from subluminal to superluminal velocities is needed for the complete residual wave (composed of both subluminal and superluminal components) to form in the smallest possible transitional region so that the light curves of both components match at their peaks. We note that many observed residual waves match at intensities other than their peak intensities, suggesting that overlap might take place. We also note that time compression effects might also result from the acceleration of the impactor wave but are beyond the scope of this manuscript.

The two emitting regions must thus each be small so that the impactor wave produce smeared-out residual light curves while these two emitting regions must be located close together to minimize the amount of overlap between them. We refer to these two conditions together as the {\em inversion simultaneity constraints}.

The near simultaneous flipping of the residual light curve might at first seem unphysical for a wave transitioning abruptly from subluminal to superluminal velocities. One would expect to see a gradually-stretched signature indicative of acceleration in the residuals near the time of reflection, but this is not observed. What is occasionally observed is a misalignment of the low-amplitude wings of the folded residuals. Strangely, this makes sense, as superluminal motion reverses the order of the low-amplitude wings and the high-amplitude central peak. Acceleration should thus cause the wings to exhibit less of a velocity stretching than the central peak. Non-linear stretching (which might indicate acceleration of the impactor wave) is observed in the wings of a few bright GRB light curves such as BATSE pulse 249 (see Figure \ref{fig:249}). However, this acceleration effect is hard to verify since it occurs in the low signal-to-noise ratio residual wings rather than in the high signal-to-noise ratio residuals near the pulse peak. We note that abruptly transitioning energy release is observed but not yet explained in other high energy astrophysical phenomena, such as in solar flare magnetic reconnection. 


The stretching factor $s_{\rm mirror}$ measured from GRB pulse residuals is the best linear fit obtained by folding the time-forward part of the residual wave and stretching it until it matches the time-reversed part of the wave. Historically \citep{hak14, hak15, hak18a, hak18b} the fitted stretching factor $s$ (and subsequently its model-independent counterpart $s_{\rm mirror}$) was inconveniently defined to be a {\em compression} of the residuals duration following the time of reflection relative to the residuals duration preceding it. In other words, small $s_{\rm mirror}$ and $s$ values refer to residual waves that need to be stretched significantly preceding the time of reflection in order to match the component following the time of reflection. For example, $s_{\rm mirror}=0.1$ indicates that the residuals following the time of reflection are 10 times longer than the time-inverted residuals preceding it. Small stretching factors ($s_{\rm mirror} \approx 0$) are found in asymmetric pulses ($\kappa \approx 1$, where $\kappa$ is the asymmetry factor of pulses fit by the \cite{nor05} function), whereas large stretching factors ($s_{\rm mirror} \approx 1$) are found in symmetric pulses ($\kappa \approx 0$).

In the acceleration model, $s_{\rm mirror}$ is the best-fit ratio of the time-reversed superluminal part of the residual light curve to the subluminal part of the residual light curve. The time intervals during which each set of residuals can be observed can be found from $\Delta t = R_{\rm wave}/u$, and $s_{\rm mirror}$ is given by


\begin{equation} \label{eqn:s}
s_{\rm mirror} = -\frac{\Delta t_i}{\Delta t_o} = -\frac{u_o R_{\rm wave}}{u_i R_{\rm wave}} =  -\frac{u_o}{u_i} = \frac{v_o (v_i/c_J - 1)}{v_i (1 - v_o/c_J)}= \frac{\beta_o [\beta_i/(c_J/c)-1]}{\beta_i [1-\beta_o/(c_J/c)]}.
\end{equation}

Equation \ref{eqn:s} can be described by three physically meaningful variables: 1) the subluminal Lorentz factor $\Gamma_o = 1/\sqrt{(1-\beta_o^2)}$ (where $\beta_o = v_o/c$), 2) the superluminal Lorentz factor $\Gamma_i = 1/\sqrt{(1-\beta_i^2)}$ (where $\beta_i = v_i/c$), and 3) the ratio of the speed of light in the jet to the speed of light in vacuum $c_J/c$.

\begin{figure}
\plottwo{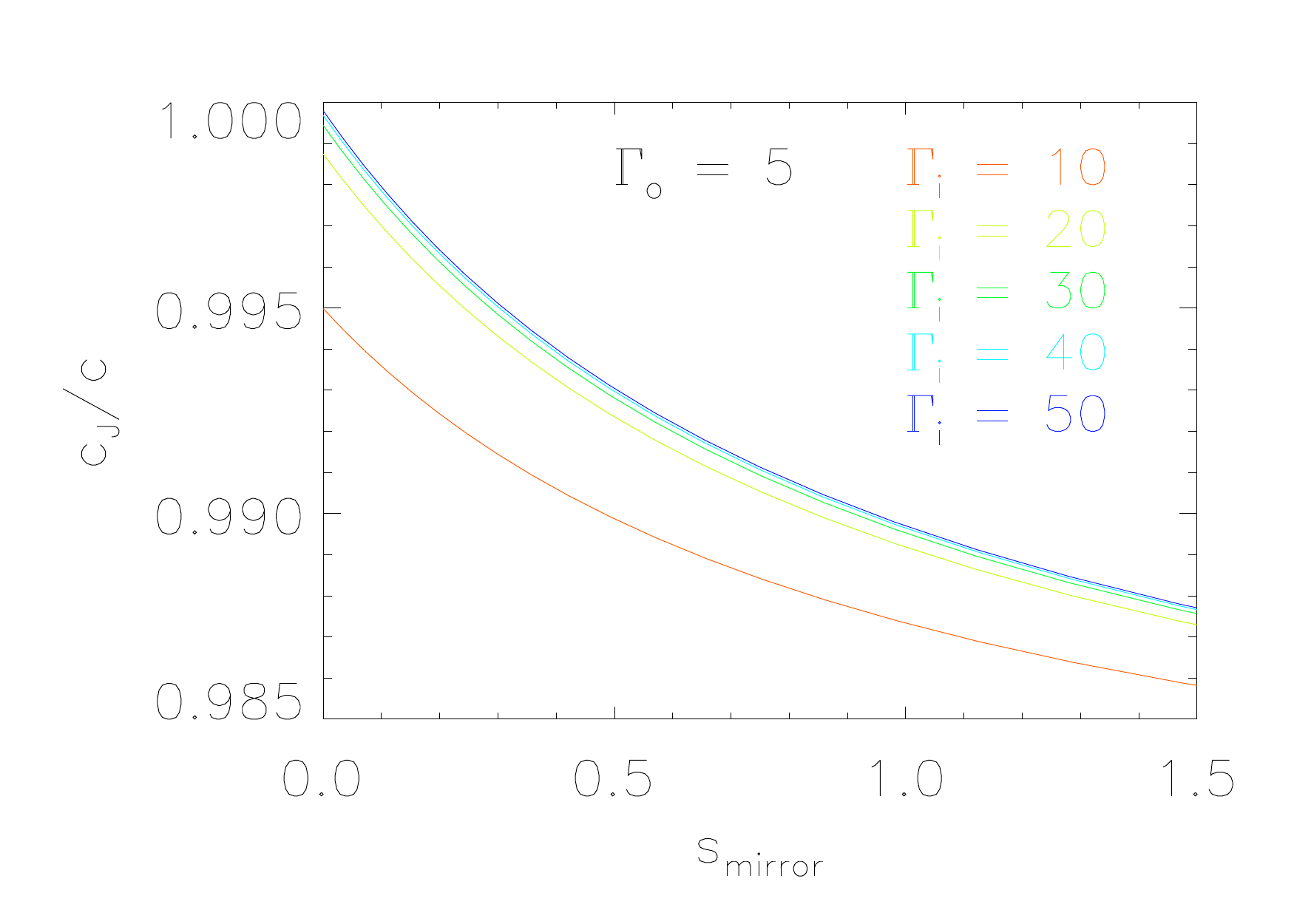}{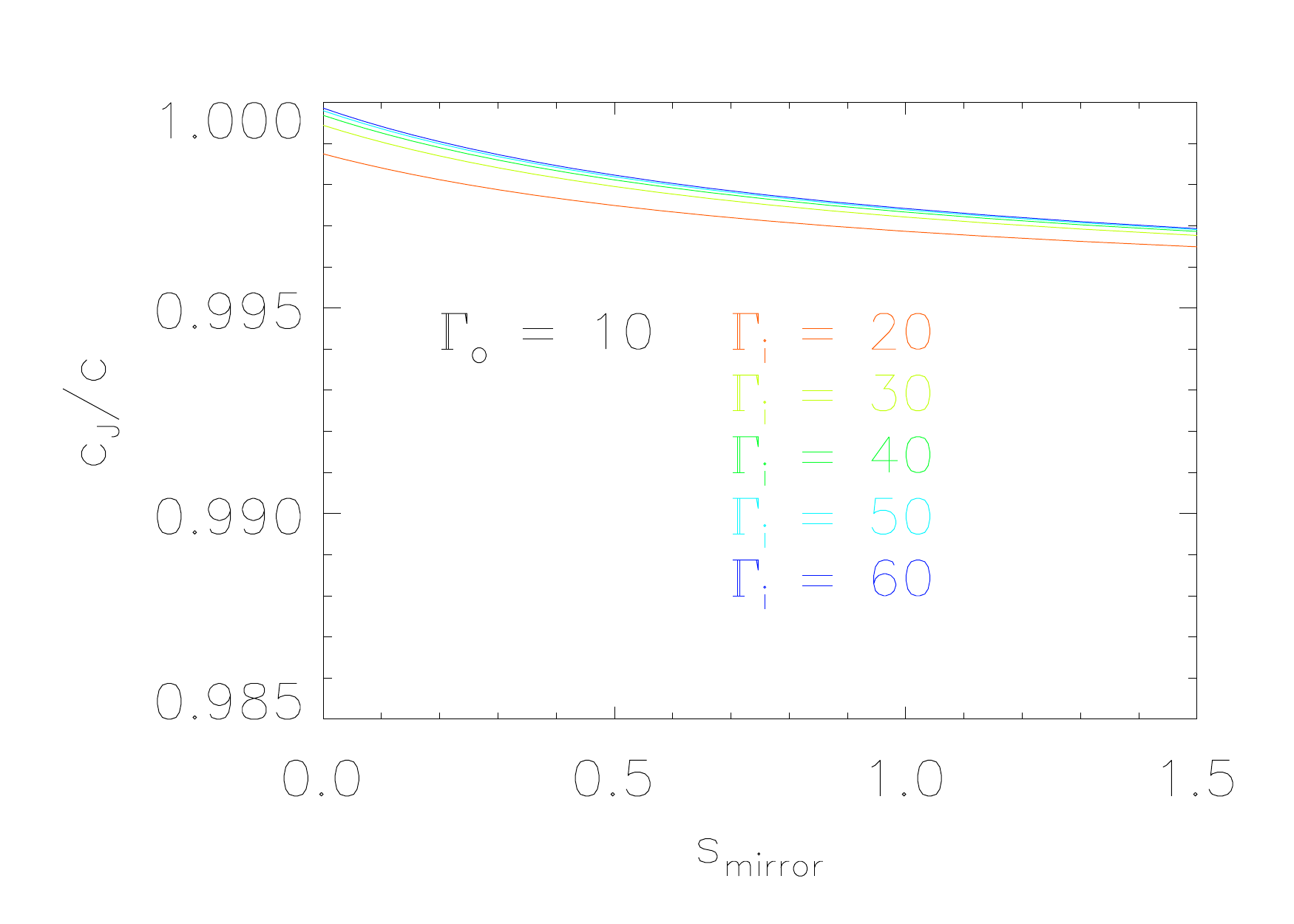}
\caption{Predicted values of $c_J/c$ for pre-accelerated impactor wave Lorentz factor $\Gamma_o$, post-accelerated impactor wave Lorentz factor $\Gamma_i$, and measured value of $s_{\rm mirror}$. A pre-accelerated impactor wave Lorentz factor $\Gamma_o=5$ is shown in the left panel. For comparison, a pre-accelerated impactor wave Lorentz factor $\Gamma_o=10$ is shown in the right panel.  \label{fig:s}}\end{figure}

Figure \ref{fig:s} demonstrates the relationship between $s_{\rm mirror}$, $\Gamma_o$, $\Gamma_i$, and $c_J/c$. The expected range of $0 \le s_{\rm mirror} \le 1$ values can be recovered for any choices of $\Gamma_o$ and $\Gamma_i$, as $c_J$ is constrained to lie in the range $v_o \le c_J \le v_i$. The speed of light values found in asymmetric pulse geometries ({\em e.g.,} those with $\kappa << 1$ and $s_{\rm mirror} \approx 1$) appear to be larger than those found in symmetric pulse geometries ({\em e.g.,} those with $\kappa \approx 1$ and $s_{\rm mirror} <<1$). This makes sense if asymmetric pulses are produced in regions characterized by lower densities or stronger magnetic fields; either of which can help facilitate photon transport. The smallest permitted values of the medium's speed of light occur when the impactor wave accelerates gradually, and larger accelerations are favored when pulses are symmetric pulses ($s_{\rm mirror} >> 0$). 

The kinematic model described here makes a number of testable predictions. Some of these relate to GRB pulse temporal properties while other relate to GRB spectral properties.

\subsection{Pulse Temporal Properties} \label{sec:time}

A key component of the RID interpretation is validating the relationship between time-reversed residual stretching and monotonic pulse shape. Three methods have been used thus far to characterize this relationship: the first combines the residuals of pulses having normalized durations, the second fits GRB pulse light curves with two independent functions (one monotonic and the other non-monotonic), and the third folds and stretches the rise portion of a pulse's residuals mathematically until it matches the pulse decay portions.

The first method was developed in the process of identifying a ``standard" GRB pulse shape. \cite{hak14} fitted 54 smooth isolated BATSE and Fermi GBM GRB pulses of moderate signal-to-noise ratio ($S/N$) with the \cite{nor05} monotonic pulse shape, subtracted each fitted pulse from the recorded counts to obtain residuals, and summed these residuals (scaled to a normalized pulse duration and intensity) to obtain a mean residual light curve. If the residuals had been distributed randomly across the pulse duration (indicative of noise), then they would have averaged to zero and the \cite{nor05} model could have been regarded as a satisfactory pulse shape. However, the resulting distribution was found to contain a wavelike feature consisting of three peaks and two valleys: the brightest peak (the {\em central peak}) occuring near the fitted pulse peak, another peak (the {\em precursor peak}) occurs at or before the pulse rise, and a third peak (the {\em decay peak}) occurs during the pulse decay. Many pulse light curves are best described by a combination of the \cite{nor05} monotonic pulse function and this residual wave, and the bumps in the residual wave often show up as additional peaks in the pulse light curve.  

The second method involved the development and application of an empirical residual fitting model. The \cite{hak14} residual wave function contains only four free parameters (one of these is the stretching factor $s$) and takes advantage of observed temporal symmetries, although the {\em asymmetry} of the underlying pulse $\kappa$ is not one of the free model parameters. The residuals are fitted independently from the monotonic pulse model. Fit improvement is identified by a $\Delta \chi^2$ test, and pulse {\em complexity} is characterized using the final goodness-of-fit as well as the $\Delta \chi^2$ improvement as a data mining classification attribute. Pulses are {\em simple} if they are best-described by a monotonic function alone, {\em blended} if they also require the residual wave model, {\em structured} if the pulse plus residual model is generally a good fit, and {\em complex} if the pulse is too structured to be captured by this model.

The third method was developed to analyze structured and complex pulses not adequately explained by the residual model. \cite{hak18b} showed via simulation that the complex structures in bright GRB pulses can take on the triple-peaked form as $S/N$ is reduced, and that the extracted properties of the fitted residual function remain constant so long as the residual structure can be identified. In effect, this analysis demonstrated that background variance dominates observations of faint GRB pulses compared to those of bright pulses. The study also showed that high-$S/N$ pulses are not limited to three peaks, but can contain an extended wave having seven peaks or more, as well as rapidly-varying spiky structures that also often appear time-reversed. Given the inadequacy of the \cite{hak14} residual function in modeling these features, a model-independent cross-correlation approach was developed to test time-reversibility: the residual structure prior to the {\em time of reflection} was folded, stretched by an amount $s_{\rm mirror}$, and tested against that following the time of reflection using a Spearman rank-order correlation test. Significant correlations were found between the time-reversed and stretched features of six bright ($S/N \approx 100$) BATSE pulses and the features observed during the pulse decay. 

The latter two methods show that the $s$ and $s_{\rm mirror}$ values of time-reversible residual structures anti-correlate with asymmetry $\kappa$ \citep{hak18b}, indicating that pulse asymmetry corresponds with an asymmetry in the residual structure. In other words, the residuals in part define a pulse's characteristics, even though they are not always associated with the commonly-recognized ``pulse." It is worthwhile analyzing a larger dataset of bright GRB pulses to test the robustness of this relationship.

We have increased this database to 31 bright GRB pulses (30 from BATSE, one from GBM) that include the six described previously. Application of the folding method produces the results shown in Figure \ref{fig:mirror}. A Spearman Rank Order correlation test between $\kappa$ and $s_{\rm mirror}$ yields a coefficient of $-0.755$ and a corresponding $p-$value of $9.3 \times 10^{-7}$. 

We have run a series of simulations to verify that pulse property extraction in a noisy background does not play a large part in recovering this strong anti-correlation. Using the best-fit pulse and residual wave model properties of each pulse, we ``noisify" these by adding a Poisson background to each to create a simulated dataset. Re-fitting both the pulse and residual properties for each pulse produces 31 $\kappa$ and $s$ values. For each of 100 Monte Carlo runs, these sample values are compared to one another using a separate Spearman Rank Order correlation test. The resulting variations produce a mean coefficient of $-0.798$ with a standard deviation of $0.010$. This corresponds to a mean $p-$value of $7.5 \times 10^{-8}$ with no simulations having as large a value as that found from the actual dataset. In other words, the pulses in Figure 5 are bright enough, complex enough, and have sufficiently-recognizable residual patterns that background noise contributes little to the fits and to the overall $\kappa$ vs.~$s_{\rm mirror}$ anti-correlation. 

Mis-identification of the pulse and the underlying residual structure can produce larger variations in the results than does noise. Using the same approach as \cite{hak18b}, we characterize the strength of the $s_{\rm mirror}$ measurements by comparing the folded and stretched residuals preceding the time of reflection with those following it using the Spearman Rank Order correlation $p-$value. All 31 pulses show strong correlations ($p < 10^{-8}$). Even when comparisons are only made between the brightest residual peaks ($p_{\rm bright}$, representing correlations between structures exceeding $3 \sigma$ from the pulse plus background fit), 25 of the 31 pulses have $p_{\rm bright} < 10^{-3}$. The six remaining pulses still exhibit excellent correlations, but constitute blended pulses in low $S/N$ data characterized by a good pulse plus residual fits not having a sufficient number of points from which $p_{\rm bright}$ can be measured.

These results support the claim that the residuals are linked chains of time-reversed events related to the underlying pulse shape that can be used to define GRB pulse characteristics. These relationships have been found in most very bright GRB pulses where signal-to-noise ratios do not smear them out: at least $80\%$ of pulses studied to date have been found to exhibit recognizable time-reversed residuals while the light curves of the remaining $20\%$ are presently too complex to interpret. 

\subsubsection{The Monotonic Pulse Component} \label{sec:monotonic}

The strong independence of the pulse shape on the residual stretching indicates whatever time-reverses and stretches the residual structure also affects the monotonic pulse structure. Also, pulses undergo generally hard-to-soft evolution. We consider three kinematic possibilities as to how the monotonic pulse might be produced to match these characteristics:
\begin{description}
\item [Option 1] The monotonic pulsed emission, like the residual emission, might be reversed and stretched at the time of reflection. This would happen if the pulse decay is a superluminal counterpart of the subluminal pulse rise. The stretching found in the decay portion of the residuals would then correspond directly to the monotonic pulse decay tail that characterizes FRED-like (Fast Rise Exponential Decay) GRB pulse light curves. However, the radiation mechanisms that produce the two parts of the monotonic pulse emission would be different, so it is hard to see how the two components could smoothly overlap to produce the observed hard-to-soft spectral evolution without being accompanied by an abrupt hardness transition in the pulse spectrum at the time of reflection.
\item [Option 2] The monotonic pulsed emission might be produced entirely during the subluminal phase. The process producing this component would continue to occur after the time the residual wave accelerates to superluminal velocities. In this case the radiation mechanism producing the monotonic pulse component would be entirely a subluminal process, with no superluminal counterpart. This process could more naturally explain hard-to-soft monotonic pulsed emission than a folded process like Option 1, but it would also have difficulty explaining the correlation between the residual stretching and the pulse shape stretching if these were produced independently.
\item [Option 3] The majority of the monotonic pulsed emission might be produced subluminally (as in Option 2), but some part of it could be produced superluminally (as in Option 1). If, for example, part of the long monotonic pulse tail is soft emission produced superluminally at the time that the folded residuals are produced, then this superluminal part could be superposed with underlying monotonic subluminal pulsed emission to produce a pulse that exhibits both hard-to-soft evolution and stretched residuals that correlate with monotonic pulse stretching.
\end{description}

Options 2 and 3 hypothesize that the residuals and monotonic pulse components are independent. In support of this, we point out that the spectral re-hardening at each residual peak might not be a characteristic of the monotonic pulse spectrum; re-hardening might simply become more pronounced when the residuals contribute more photons to the overall spectrum, and this re-brightening might simply disrupt the graduate spectral evolution of the monotonic pulse component. 

\subsubsection{Stretch Factors Exceeding Unity} \label{sec:bigstretch}

\begin{figure}
\epsscale{0.50}
\plotone{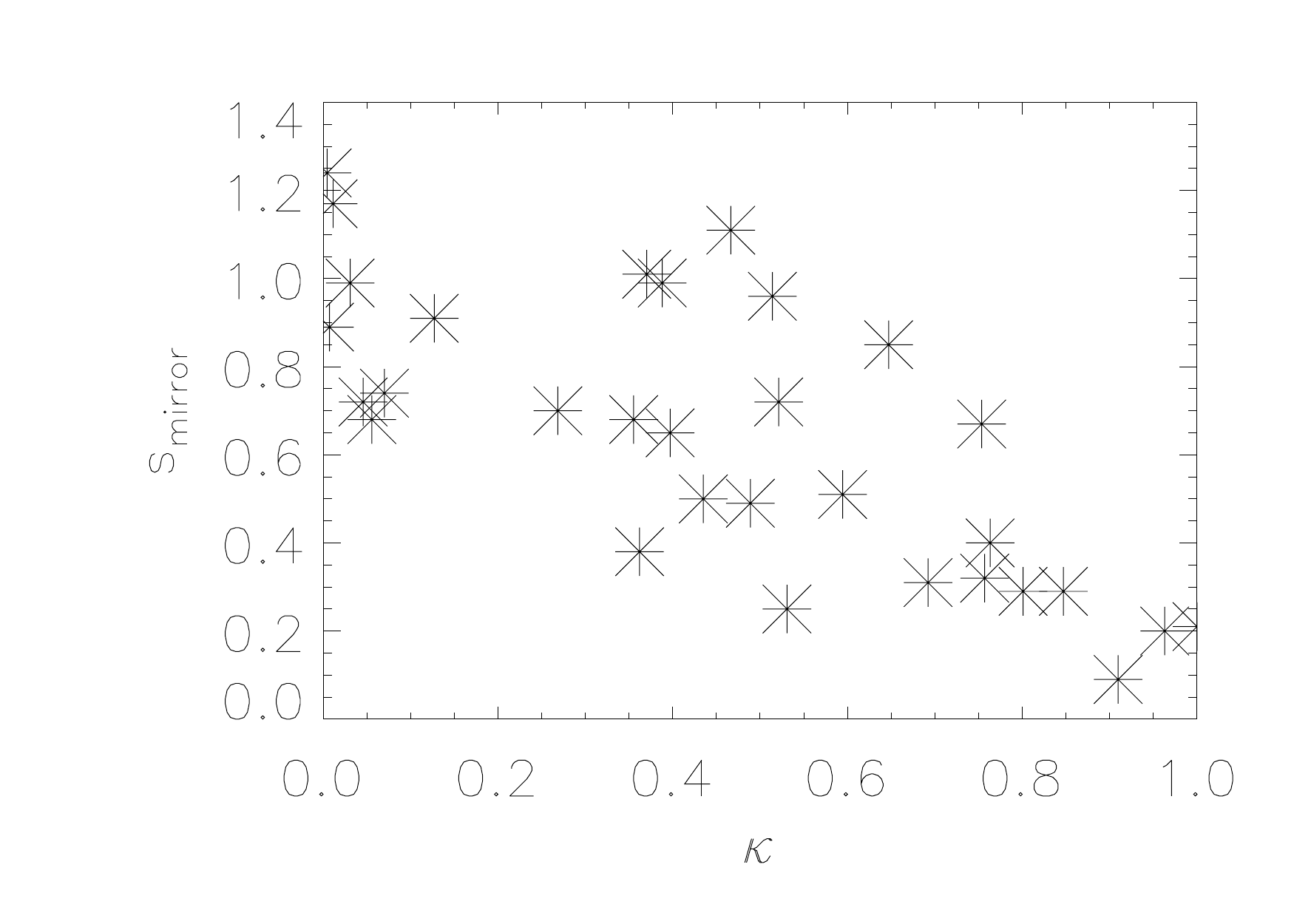}
\caption{Observed values of $s_{\rm mirror}$ vs.~ $\kappa$ for 31 GRB pulses. Note that some $s_{\rm mirror} > 1$ are measured, which is predicted for some jets transitioning from subluminal to superluminal velocities. A Spearman Rank Order correlation test finds an anti-correlation of -0.755 between the parameters, for a $p-$value of $9.3 \times 10^{-7}$. \label{fig:mirror}}\end{figure}

As can be seen in Figure \ref{fig:s}, the Impactor Acceleration model predicts $s_{\rm mirror} > 1$ for larger values of $c_J/c$. This is unexpected since the mathematical form chosen for the monotonic pulse model \citep{nor05} does not allow for asymmetric pulses in which the rise time exceeds the decay time ({\em e.g.,} $\kappa > 1$).  However, of the 30 bright GRB pulses exhibiting time-reversed residuals, (see Figure \ref{fig:mirror}) four require fits in which $s_{\rm mirror} > 1$. The residuals of these pulse residuals are asymmetric with longer-duration structures occurring during the pulse rise than during the decay.  Because pulses with $s_{\rm mirror} > 1$ exist, we cannot eliminate the possibility that even larger values of $s_{\rm mirror}$ might be present in GRB pulses. Very large $s_{\rm mirror}$ values would have the consequence of making it very hard to uniquely identify pulses by their time-reverse and stretched residuals. Therefore, we hypothesize that at least some of the bright pulses in which time-reversed residuals cannot be identified might simply result from very large $s_{\rm mirror}$ values that could make their time-reversible natures unrecognizable.

\subsubsection{Timing Issues Relating the Prompt Emission to the Afterglow} \label{sec:afterglow}

Inversion simultaneity requires that the region containing the transition from subluminal to superluminal motion must be at least the radial size of the impactor wave ($\Delta R \ge R_{\rm wave}$); this seems easiest to generalize when $\Delta R = R_{\rm wave}$. This condition seems remarkably fortuitous unless the beginning of the wave impactor is responsible for creating the superluminal emission conditions and the end of the wave impactor is responsible for terminating the subluminal emission conditions, and the two events occur nearly simultaneously. We do not know how such conditions might arise and be temporally linked, but if this is the case then it is suggestive of conditions that exist within both strong magnetic fields and relativistic shocks. The inversion simultaneity constraint might not be so rigid if some of our assumptions turn out to not be correct. For example, this condition might be relaxed if the speed of light in the medium $c_J$ is not constant. Similarly, it might be relaxed it the impactor wave produces a gradient in the medium's index of refraction as it passes through it and produces radiation. Regardless, we view the simultaneity constraints as critically important yet difficult-to-explain parts of this model.

Separate subluminal and superluminal emissions create a temporally-ordered light chain that follows in the wake of the superluminal impactor wave. RID requires that the impactor wave must slow from superluminal to subluminal velocities before the emission produced during its superluminal phase can be observed \citep{nem18}. Thus, the time-reversed part of the light curve does not have to be attached to the light chain when the impactor wave is close to the progenitor (where the forward part of the light curve is produced), but can wait to be attached until the impactor wave slows closer to the region in which the afterglow is created. Regardless, the 'time-forward' and 'time-reversed' parts of the light chain will appear to happen almost simultaneously. This coupling might also explain how the very high energy emission usually associated with the afterglow can appear to start at almost the same time as the prompt emission ({\em e.g.,} see the review by \cite{pir16}): the superluminal impactor can slow and plow into the external medium while simultaneously setting up the conditions where the prompt emission can catch up with it.

\subsubsection{Multi-Pulsed GRBs} \label{sec:multi}

Each pulse memorializes the transition of an impactor wave from subluminal motion to superluminal motion within a GRB jet.
Thus, one of two conditions is necessary for producing a multi-pulsed GRB: 1) the central engine only ejects material once, in which case a subsequent pulse results from a second acceleration of this material from subluminal to superluminal velocities, or 2) the central engine ejects material more than once, in which case a subsequent pulse occurs only when the newly-ejected material is also accelerated from subluminal to superluminal velocities. 

\begin{description}
\item [Case 1] A progenitor only ejects a single impactor wave. This is expected from a cataclysmic explosion. In this case, the single impactor wave is accelerated from subluminal to superluminal velocities repeatedly in a GRB jet. The process by which material is made to move faster than the speed of light in a medium is an energetically-depleting one \citep{tam60}. Thus, it is possible that jet expansion does not smoothly accelerate the impactor wave once, but might sometimes be required do so repeatedly with some energy loss each time until superluminal velocities have been achieved (this is somewhat analogous to skipping a stone off of the surface of a pond). In other words, the impactor wave might be briefly accelerated beyond $c_J$ until it has radiated away enough energy to become subluminal. However, expansion causes acceleration to occur again, producing additional pulses to form.
\item [Case 2] A progenitor can eject multiple impactor waves. This might be expected if an accretion disk only ejects some of the mass available to it in an explosion, after which it must wait until additional mass infall produces these conditions again. 
\end{description}

GRB pulse durations increase as interpulse separations increase in short GRBs, long/intermediate GRBs \citep{hak18a, hak18b, pat18}, and x-ray flares \citep{mar10}.  Using the old GRB pulse definition where a pulse was any intensity fluctuation not explained by noise, durations of {\em emission episodes} were found to increase as bursts progress \citep{rrm01a,rrm01b,dp07}, whereas ``pulse" durations remained constant \citep{rrm00}). Our results are consistent with these observations, although the revised pulse definition (our ``pulses" are their ``emission episodes" and our ``structured variations" are their ``pulses") argues against the need for internal shocks and replaces these with impactor wave emission. Thus, if the central engine ejects material only once, then the increased pulse duration associated with delayed pulse emission suggests that the jet ejecta loses kinematic energy and thus velocity as it moves outward. If the central engine can eject material repeatedly, then the time interval between pulses indicates the time it takes for the central engine to recharge while the corresponding pulse duration increase suggests that subsequent pulses are moving slower than their early counterparts.

\subsection{Pulse Spectral Properties} \label{sec:spectra}

Cherenkov radiation \citep{che34,tam37} can be emitted when an impactor's velocity exceeds the speed of light in the medium through which it travels; this radiation can be produced when an impactor travels at constant velocity and does not require the emitting particles to be accelerated. Cherenkov radiation is a continuous form of radiation that allows ultra-relativistic electrons and ions to draw energy from plasma waves as they coherently ``ride the waves"
similar to how shock waves are produced by aircraft exceeding soundspeed. This energy extraction from plasma waves is not the only mechanism by which an impactor can produce energy superluminally; particle collisions capable of producing potentially significant numbers of photons can still occur \citep{fer40} within the Debye radius $D=(\epsilon_0 kT/ N e^2)^{1/2}$ ({\em e.g.,} \cite{tam60}; where $\epsilon_0$ is the permittivity of free space, $T$ is the plasma electron temperature, $N$ is the electron density, $e$ is the electrostatic charge, and $k$ is Boltzmann's constant).  The number of interactions that occur is critical to the production of Cherenkov radiation. As a result this radiation is observed on Earth primarily in high-density regimes such as when cosmic rays or accelerated particles pass through solids and/or liquids. In a relativistic GRB plasma, the production of Cherenkov photons would be expected to occur in regions where a high density of superluminal particles interact over a large path length containing a high density of target particles. Although a GRB jet contains a large number of superluminal particles traveling over a long path length and magnetic fields that can align and collimate the particles, it is not known if the Cherenkov process alone might produce photons in the quantities needed match the intensities of time-reversed GRB pulse light curves, or if collisions might provide a significant contribution to the flux. It is also not known whether the Cherenkov process is capable of absorbing photons in a way that is consistent with the observed residual wave structure, since this appears to extract photons from the monotonic pulse light curve. 

Although Cherenkov emission is emitted isotropically, it is brightest when observed at an angle $\theta = \cos^{-1}(c_J/v)$ with respect to the direction of motion.
This angle constitutes the longest stretch of radiation per second triggered by the superluminal particles that is seen by the observer.

Cherenkov spectra increase in intensity at higher energies (this gives Cherenkov spectra on Earth a characteristic violet glow), but are cut off in an unmagnetized plasma at x-ray and $\gamma-$ray frequencies with photon energies greater than the plasma frequency $E=e^2 N/(\epsilon_0 m_e)$ (where $m_e$ is the electron mass) because the index of refraction, being wavelength-dependent, is less than unity. 
The presence of a magnetic field can make Cherenkov radiation possible in a plasma ({\em e.g.,} \cite{sol65}), but the relative strength of Cherenkov radiation to synchrotron radiation depends on the velocity of the material and other physical conditions within the plasma. Conventional relativistic magnetohydrodynamics (MHD) is generally used to model relativistic magnetized plasmas ({\em e.g.,} \cite{ani90}). However, this approach fails if the timescale defined by the Larmor radius is shorter than the thermal collision time \cite{ten08}, in which case Lorentz-covariant, general relativistic fluid equations and a stress tensor that accounts for pressure anisotropy and heat flow parallel to the magnetic field are preferred. A treatment such as this reduces to the standard MHD solutions in the non-relativistic and mildly-relativistic limits, but demonstrates that the phase velocities of a variety of plasma modes remain subluminal in the relativistic limit for a single-species plasma. Results like this might be necessary for predicting the characteristics of Cherenkov emissions from a coherent superluminal impactor wave. The composition of the plasma might also be important; \cite{sco04} and \cite{fil05} have shown that indices of refraction greater than unity can develop in plasmas composed of certain ionic species.

In order for the Impactor Acceleration model to be consistent with GRB pulse spectral evolution, early subluminal emission must be spectrally harder than the later superluminal spectrum. This requires the radiation produced during the superluminal phase to be softer than that emitted during the subluminal phase. Since the mechanistic details that produce the subluminal emission are still unknown, this condition is difficult to verify at present time. Complicating this is the fact that the form and energy of the suspected superluminal spectrum is also unknown; especially since it likely depends on free parameters such as the mass distributions of relativistic particles in impactor waves, the conditions that cause the impactor wave to become superluminal, the local magnetic field, the jet geometry, and initial and final impactor wave Lorentz factors.

One observed characteristic that the Impactor Acceleration model might natural explain is the appearance of thermal spectral components during early emission of some GRB pulses. A blackbody-like spectrum might be produced in a region of high density such as where the impactor wave expands at $c_J$ (the boundary between subluminal and superluminal motion) or perhaps close to the stellar surface.

\section{Impactor Wave Deceleration} \label{sec:alternate}

\begin{figure}
\plotone{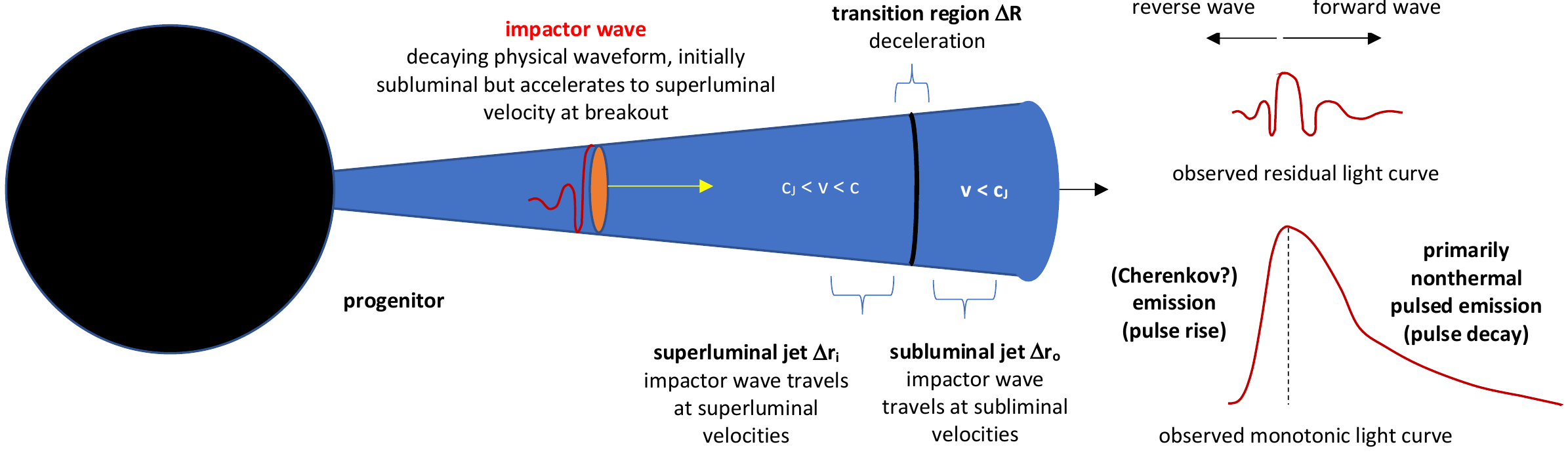}
\caption{The Impactor Deceleration Model. Within a GRB jet, an impactor wave accelerates from subluminal to superluminal velocities in the outer part of the jet. In contrast to the Impactor Acceleration model, the impactor wave starts at its maximum value, and gradually tapers off.  \label{fig:dmodel}}\end{figure}

Equally viable to Impactor Acceleration is a model in which RID might create time-reversed GRB light curves when an impactor wave decelerates from superluminal to subluminal velocities. This can be visualized by changing the order of the light spheres in Figure \ref{fig:accelerate}. The model for this deceleration (Figure \ref{fig:dmodel}) is similar to that in Figure \ref{fig:model}, but would occur farther out in the jet, when the impactor decelerates prior to the onset of the afterglow phase. 

Some aspects of this model are more consistent with the standard GRB paradigm than the Impactor Acceleration model, since all GRB shock models generate light from the dissipation of kinematic energy ({\em e.g.,} \cite{ree92,pan98,der99a,der99b,gib02}), rather than from adiabatic expansion. However, the Impactor Deceleration model differs from the standard external shock model in that the energy generation process occurs from an impactor moving faster than the speed of light rather than from one moving faster than the speed of sound.

The Impactor Deceleration model requires the first part of the residuals to be a time-reversed component produced when the impactor wave is traveling at superluminal speeds, and second part to be produced when the impactor wave ploughs into an external medium. In order to produce observed GRB pulse light curves, the impactor wave must have a temporally inverted form from that shown in the Impactor Acceleration model (compare Figure \ref{fig:model} to Figure \ref{fig:dmodel}). In other words, the impactor waveform must begin with an initial high-amplitude spike which is then followed by a decaying waveform. Physically, this structure might represent something like an explosive outflow at the head of the jet followed by decaying instabilities within the jet.

The inversion simultaneity constraint again indicates that the emitting regions have small radial extents. These two regions can only be close together if the superluminal impactor velocity $v_i$ is much greater than the speed of light in the medium. In other words, if $v_i >> c_J$ then $\Delta R \approx R_{\rm wave}$.

Due to symmetry, Equation \ref{eqn:s} still describes the amount of stretching observed in the time-reversed wave  (if the speed of light in the medium $c_J$ remains constant during the transition), with the simple caveat that and $c_J  < v_o < c$ (the initial velocity is superluminal) and $v_i < c_J < c$ (the final velocity is subluminal). It seems peculiar that the same equation describes time-reversed stretching for both acceleration and deceleration, but replacing subluminal with superluminal velocities causes the symmetric equation to produce asymmetric results. In effect, deceleration solutions describe conditions when the light wave separations in subluminal light spheres are larger than those in time-reversed superluminal light spheres, instead of the other way around.

\begin{figure}
\plottwo{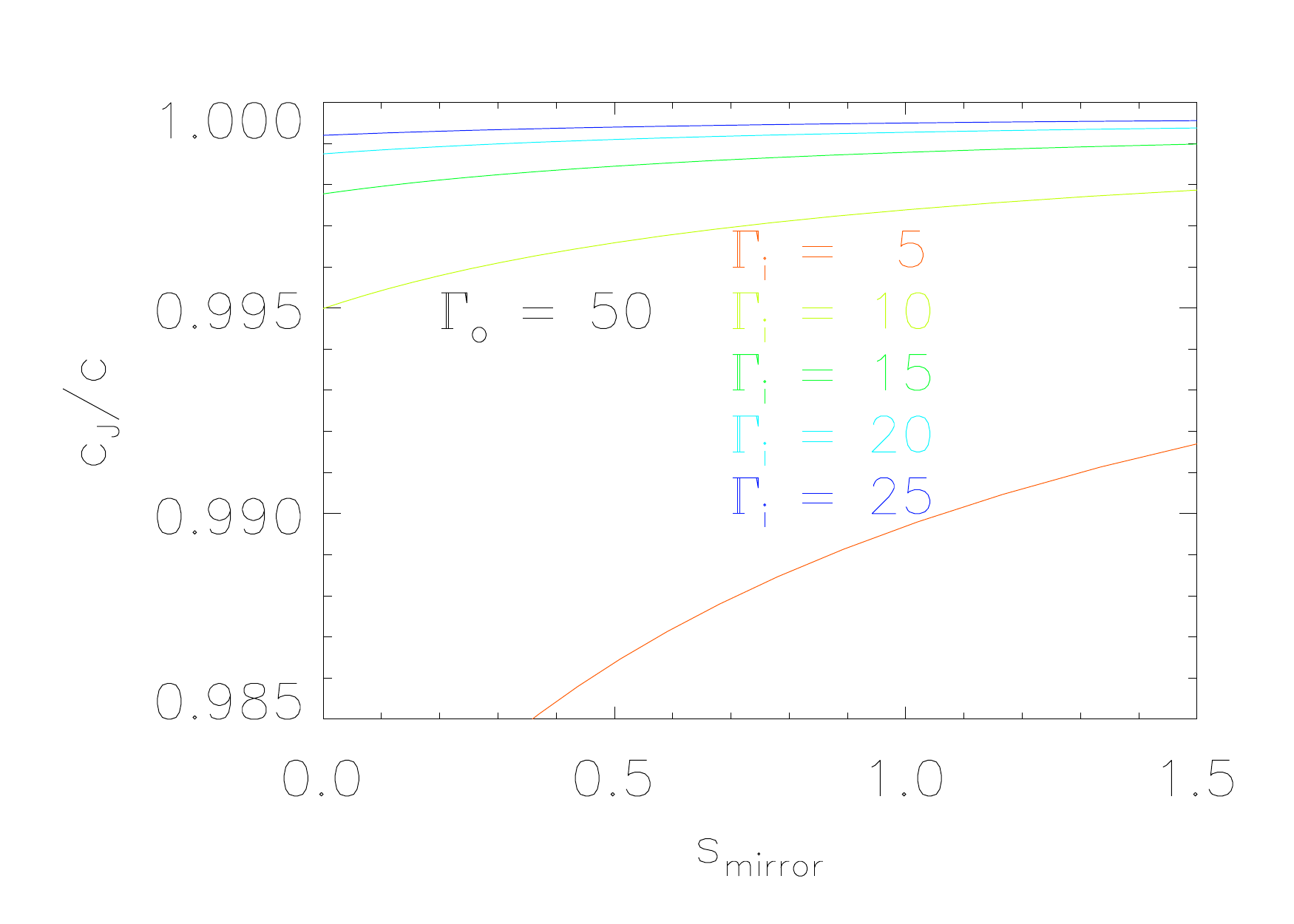}{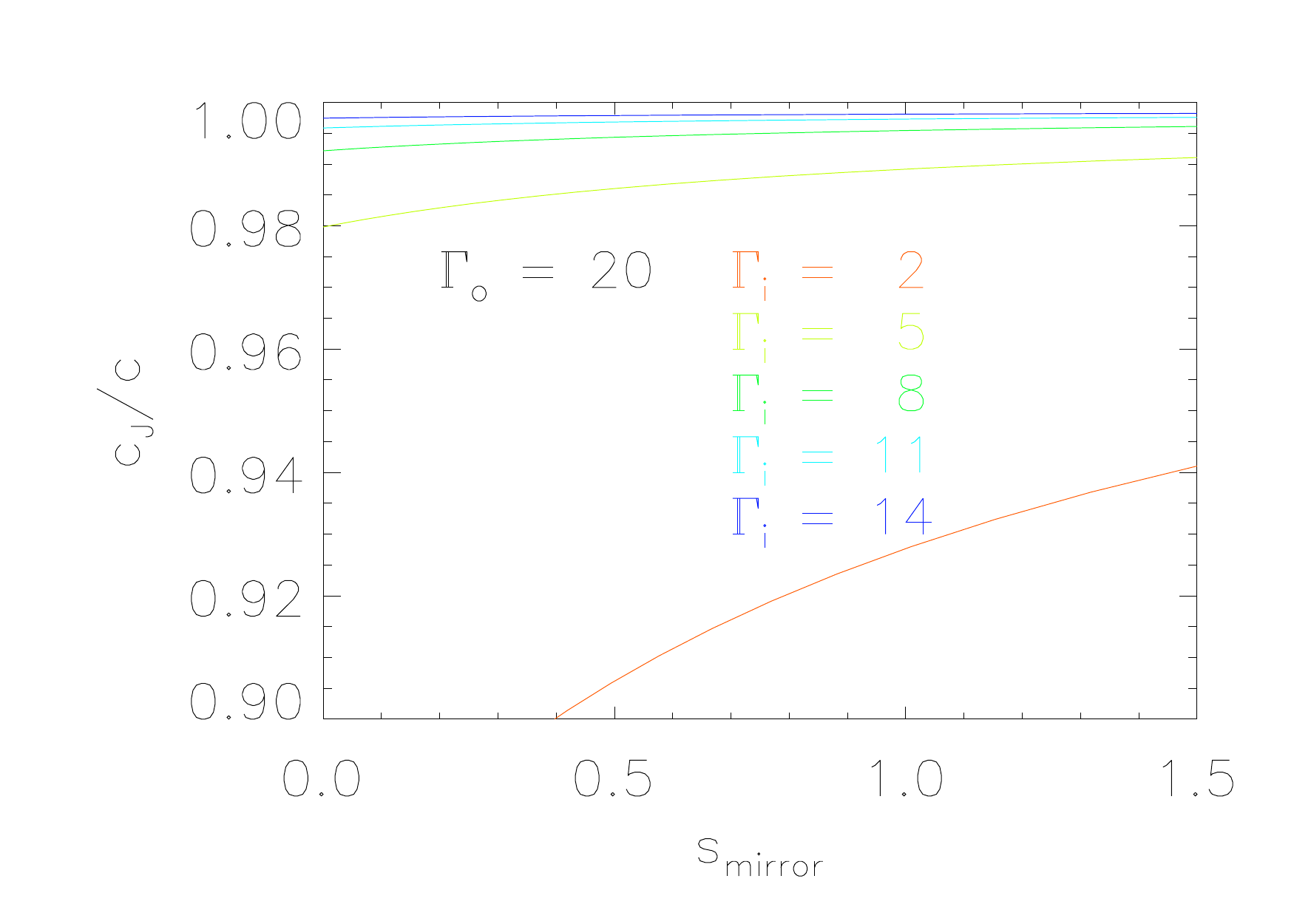}
\caption{Predicted values of $c_J/c$ for pre-decelerated impactor wave Lorentz factor $\Gamma_o$, post-decelerated impactor wave Lorentz factor $\Gamma_i$, and measured value of $s_{\rm mirror}$. A pre-decelerated impactor wave Lorentz factor $\Gamma_o=50$ is shown in the left panel. For comparison, a pre-decelerated impactor wave Lorentz factor $\Gamma_o=20$ is shown in the right panel.  \label{fig:sd}}\end{figure}

The resulting relationships between $\Gamma_o$, $\Gamma_i$, $c_J$, and $s_{\rm mirror}$ are shown in Figure \ref{fig:sd}. These are similar to those predicted for the Impactor Acceleration model shown in Figure \ref{fig:s} in that $c_J$ is again constrained to lie in the range $v_o \le c_J \le v_i$.  The smallest permitted values of the medium's speed of light occur when the impactor wave decelerates suddenly, and greater decelerations are favored when pulses are asymmetric pulses ($s_{\rm mirror} << 1$). In contrast to the Impactor Acceleration model, this only makes sense if asymmetric pulses are produced in regions characterized by higher densities or weaker magnetic fields; either of which can reduce photon transport. For abrupt, large decelerations, the speed of light in the medium drops to very low values that are perhaps inconsistent with physical conditions in the medium. 

Spectrally, impactor deceleration should produce different spectra from the impactor acceleration.  For the Impactor Deceleration model, the superluminal (Cherenkov) component produced by superluminal motion must be produced prior to the subluminal component and must be spectrally harder. That is the opposite of the Impactor Acceleration model, where the superluminal component must be produced later and must be spectrally softer. This might be possible if Lorentz boosts and/or a locally large Debye radius permit Cherenkov radiation to be produced in the gamma-ray range. The Impactor Deceleration Model might make it harder to explain why two spectral components are observed during the pulse rise phase, as an optically-thick component is only likely to be observed during the subluminal phase. One possibility is that this component is a Cherenkov rather than a thermal component.


Like the Impactor Acceleration model, the Impactor Deceleration model predicts that a pulse can only be observed when the superluminal impactor wave drops from superluminal to subluminal velocities. As before, multiple pulses in a GRB can be caused either because the jet undergoes repeated decelerations or because the progenitor is active multiple times. The rapidity at which light carries the message of these interactions again suggests that the afterglow phase will be observed to start soon after the first prompt pulse is observed, since afterglow emission will be caused immediately after the deceleration that produced the first prompt pulse.

\subsection{Are Both Acceleration and Deceleration Possible?} \label{sec:both}

Is it possible that pulses are produced both during the impactor acceleration phase and during the impactor deceleration phase? Such a possibility suggests that GRBs should generally contain even numbers of pulses, unless geometric constraints prevent us as observers from seeing one or the other. This is clearly not the case, since single-pulsed GRBs dominate the GRB landscape. Furthermore, as indicated previously, the spectra of pulses produced during acceleration should differ from those produced during deceleration, even as their residual structures and pulse shapes should be nearly identical.

{\bf \section{Relativistic Viewing Effects}}

\subsection{Alignment of the Residuals with the Monotonic Pulse Component} \label{sec:align}

Our model has not addressed the fact that the time of reflection for pulse residuals does not always align with the peak of the monotonic pulse, although this is generally true. Figures \ref{fig:model} and \ref{fig:dmodel} are oversimplifications meant to represent generic situations. 
There is a spread of these alignments ({\em e.g.}, see Figure 19 in \cite{hak18a}), and a prime example of a pulse in which these are offset is BATSE 143p1 (see Figure 1 in \cite{hak18b}). In addition to this offset, pulse residuals are not always folded at the peak of the residual function; many pulses have time-reversed residual structures for which the time of reflection does not occur other at the maximum of the residual function (see Figures 14-16 in \cite{hak18b}). 

The offsets between the monotonic pulse peak and the residual time of reflection are partially due to instrumental response, but are also partly intrinsic. \cite{hak11} proposed that pulse lags are indicators of hard-to-soft pulse evolution, and that measured $E_{peak}$ values depend somewhat on how the intrinsic pulse $E_{peak}$ drops through the detector's response. Of course it also depends on how $E_{peak}$ is defined. \cite{pre15} demonstrated that this effect can indeed be produced in some cases by instrumental response. However, unlike the monotonic pulse components, the spectra of the residual structures do NOT appear to evolve; although the spectrum of a pulse's monotonic component generally evolves from hard-to-soft, the overall pulse spectra re-harden each time intensity peaks during residual emission rebrightening (see Figures 17-21 in \cite{hak18b}, Figures 19-28 in \cite{hak15}, and Figures 20-22 in \cite{hak15}). These intrinsic characteristics can produce observed pulse spectral evolutions that range from hard-to-soft to intensity tracking, and they can even produce negative spectral lags (see Figures 19-21 in \cite{hak18b}).

This misalignment suggests that the true situation may not be as simple and straightforward as the models proposed in Figure \ref{fig:model} and Figure \ref{fig:dmodel} imply. However, variations in alignment between the residuals and the pulse peak may provide clues to the underlying physics rather than being detriments of the model. The current models make no mention of jet orientation (is it pointing directly at us, or is it slightly off-axis?), jet geometry (does the jet flare out as it expands?), variations in stellar metallicity (what is the distribution of particles being accelerated?), magnetic field structure (are some jets more highly magnetized than others? Are pulses all produced in regions with similar magnetic field alignments?), etc. For example, the misalignment between the peak of the monotonic pulse and the rime of reflection might indicate an off-axis jet. It is encouraging that the simple models presented in Figure \ref{fig:model} and Figure \ref{fig:dmodel} can explain basic pulse properties without relying on many complications that surely must be present.

\subsection{Radiative Cooling and Jet Curvature Effects} \label{sec:curve}

\subsubsection{Radiative Cooling} \label{sec:jcool}

\cite{lia96} studied the decay portion of 37 BATSE pulses (using a structured pulse definition similar to the one we have developed) and found that $E_{\rm peak}$ (the peak in the instantaneous $\nu F_\nu$ spectrum) decayed with photon fluence at constant rates; their results were consistent with $\gamma-$radiation cooling by processes such as multiple Compton scatterings in confined plasmas containing fixed numbers of particles. The $E_{\rm peak}$ decay was found to reset each time a structured peak occurred (consistent with the observations of \cite{hak15}), and the decay constant was found to be similar for subsequent pulses in multi-pulsed GRBs. \cite{lia96} interpreted this as a mechanism by which ``the different pulses are produced by a regenerative source rather than a single catastrophic event." We note that an impactor wave provides a reasonable example of such a regenerative source. The \cite{lia96} cooling explanation is useful in describing the subluminal emission, but we do not know at this time how Cherenkov radiation produced during the superluminal phase might take on characteristics similar to cooling observed in the subluminal phase. 

\subsubsection{Jet Curvature} \label{sec:jc}

In the standard jet model, misalignment of the jet with the observer's line-of-sight causes delays between high-energy and low-energy emission. An expanding jet causes the observer to see emission coming from an increasingly off-axis annulus. This annulus contains smaller Lorentz expansion factors, coupled with geometric shell curvature, and thus produces delayed, softer radiation \citep{fen96,sal00,kum00,iok01,qin02,qin04,der04,she05,lu06}. Although models containing these features have been well-developed, they do not adequately match the observations: 1) The decay rate of a GRB pulse does not match the predictions made by curvature models ({\em e.g.} \cite{koc03,she13}). 2) Asymmetric and hard pulses are observed to undergo spectral decay from the moment they are first detected \citep{hak15}, suggesting that the curvature model should apply to the entire pulse and not just the decay portion of as is generally modeled, and 3) Curvature models do not account for bumps in the otherwise smooth light curve \citep{hak14} or the abrupt change in $E_{\rm peak}$ when these occur \citep{lia96}, much less time-reversed characteristics. Jet curvature does not appear to play a major role in creating GRB pulse shapes or in explaining their hard-to-soft spectral evolution.

\section{Conclusions} \label{sec:conclusions}

The goal of this manuscript has been to provide an explanation for the time-reversed and stretched residual light curves observed in GRB pulses; these features are not a natural byproduct of current MHD GRB models. To explain these observations, we have proposed a kinematic modification of the standard GRB model in which a wavelike impactor produces radiation as it moves both subluminally and superluminally through an expanding jet. The time-reversed light curves can be produced either when the impactor wave accelerates from subluminal to superluminal speeds or when it decelerates from superluminal to subluminal speeds. The model can explain the amounts of stretching observed in time-reversed GRB pulse residuals, the relationships between stretching factor and pulse asymmetry, and the relationships between the monotonic pulse component and the residuals. The odd numbers of pulses found in most GRBs suggest that pulses are produced by either acceleration transitions or by deceleration transitions, but not by both. Since all GRB classes exhibit pulses having similar structures, this model is just as viable for merging neutron star progenitors as it is for massive star progenitors.

Superluminal motion is unavoidable when a highly relativistic jet enters almost any medium. It happens whenever a gamma-ray photon or cosmic ray enters the Earth's atmosphere, and whenever and wherever highly relativistic particles are involved. Despite this, we do not know if a system can produce waves of radiation coherently via Cherenkov radiation. Thus, additional work is needed to determine the relative contributions of this and other scattering mechanisms. We have assumed that the impactor responsible for creating a GRB pulse must represent some large-scale wavelike physical phenomenon that might be caused by abrupt variations in density, pressure, and/or magnetic field, but exact nature of such variations is unknown; these variations might be produced within the jet or might be remnants of progenitor activity. The impactor wave must produce residual radiation in the form of a wavelike light curve on two separate occasions: once as it moves at subluminal velocities and again after it has transitioned to superluminal velocities; it is unknown why this would be. These models are not possible if plasmas cannot be found to be transparent to coherent radiation at superluminal velocities.

Is one interpretation (acceleration vs.~deceleration) preferred over the other? Since the extragalactic nature of GRBs was uncovered, it has been suggested that energy dissipation from deceleration is favored ({\em e.g.,} \cite{ree92}); it is only due to kinematics and the inefficiency of accelerating matter beyond a medium's lightspeed that we have suggested the Impactor Acceleration model as an alternative. Wave structures following the jet head are predictive features of Adaptive Mesh models ({\em e.g.,} \cite{mor07,lopez13}); these seem to support the Impactor Deceleration model, whereas models in which instabilities develop in the forming accretion disk \citep{lin16} or its magnetic fields \citep{llo16} prior to exploding favor the Impactor Acceleration model. The possible presence of thermal-like spectra during the pulse rise seems to favor the Impactor Acceleration model, unless this spectrum is what might be expected from Cherenkov radiation.  It is unclear whether Cherenkov radiation should be spectrally harder or softer than emission expected from subluminal GRB radiation, so we cannot tell which model is currently favored by pulse evolution. We note in passing that evolving polarization might be expected as Cherenkov radiation transitions to non-Cherenkov radiation (or {\em vice versa}); it is still early to tell if observations support GRB polarization evolution \citep{yon12,zha19}.

Brighter, structured GRBs are more luminous than fainter GRBs having smoother light curves ({\em e.g.,} \cite{nor00,rei01,hak08}), indicating that brighter GRB pulses are also the most luminous ones. Thus, the smoothing out of pulse structure at lower signal-to-noise ratios \citep{hak18a,hak18b} is not solely an extrinsic effect, suggesting that pulse structure is in part a characteristic of luminous pulses. One might expect luminous GRB pulses to exhibit other differences. Late pulses already exhibit longer durations, lower amplitudes, and softer spectra than early pulses, which could be an indicator of lower Lorentz factors. X-ray flares occur much later than prompt GRB pulses, have lower energies and luminosities, and are much longer than GRB pulses. Other characteristics, such as the shape of the monotonic portion of x-ray flares relative to GRB pulses, may also differ \citep{hak16}.


We have shown that RID provides a new explanation for the time-reversed and stretched residuals found in GRB pulse light curves. Despite simplifying assumptions about the geometry of the GRB jet, its composition, and the characteristics of the expanding medium, RID seems to be more consistent with other inferred characteristics of GRB jets than the bilaterally-symmetric models proposed by \cite{hak18b}. And all of these models provide better explanations for the observed GRB pulse characteristics than models which have not recognized these characteristics.


ACKNOWLEDGEMENTS: We thank the anonymous referee for helpful comments. We express our gratitude to Tim Giblin, Rob Preece, Amy Lien, and Narayanan Kuthirummal for their many patient discussions and valuable contributions to this manuscript during its development.



\end{document}